\documentstyle[12pt]{article}
\begin{document}

\hfill{\large RPI-98-N116}
 
\hfill{\large FTUV 98/20}

\hfill{\large IFIC 98/20}
\vspace{.2cm}
\begin{center}
{\large{\bf INCLUSIVE MUON CAPTURE IN LIGHT NUCLEI\\}}
\end{center}

\vspace{1cm}

\begin{center}
Nimai C. Mukhopadhyay$^{a,b}$, H. C. Chiang$^{a,c}$, S. K. Singh$^{a,d}$ and
E. Oset$^a$
\end{center}

{\small{\it $^a$Departamento de F\'{\i}sica Te\'orica and IFIC,
Centro Mixto Universidad de Valencia-CSIC,
46100 Burjassot (Valencia), Spain}}

\vspace{0.2cm}
 
{\small{\it$^{b}$ Department of Physics, Applied Physics and Astronomy,
Rensselaer Polytechnic Institute, Troy, New York 12180-3590, USA}}

\vspace{0.2cm}

{\small{\it $^c$ Institute of High Energy Physics, Academia Sinica,
Beijing, P.R. China}}

\vspace{0.2cm}

{\small{\it $^{d}$ Physics Department, Aligarh Muslim University, Aligarh,
India 202002}}

\vspace{3cm}

\begin{abstract}
{\small{We study total muon capture rates in light (A $\sim$ 6-18) nuclei,
taking into account renormalizations
of the nuclear vector and axial vector strengths.
We estimate the influence in the results 
of uncertainties of the spin-isospin
interaction parameter $g'$ and nuclear densities.
A few of these reactions are theoretical benchmarks for physics
involving searches for neutrino oscillations. New experiments in
muon capture in several targets are suggested, in the light of some
discrepancies with theory, 
crudeness of some experimental
results and relevance to neutrino physics.
}}
\end{abstract}

\newpage

Since its discovery in the forties \cite{1}, nuclear muon capture (NMC) has
been a probe {\it par excellence}
for both weak interaction studies and exploring
nuclear dynamics \cite{2}. In the context of QCD, the nucleus provides
a new vacuum, wherein quark and gluon condensates can have entirely different
values
from those in the free hadrons \cite{3}, thereby opening the possibility
of nuclear renormalization of the hadronic weak couplings \cite{4}. On the other
hand, particle-hole and/or isobar-hole excitations provide a more
traditional theoretical setting for describing the nuclear renormalization
dynamics \cite{5}. We explore here the latter in NMC via the study of inclusive
(or total) capture rates.

The importance of NMC in light nuclei (here, those with
mass number $A < 20$) transcends nuclear physics. Thus,
muon capture reactions in $^{12}C$ and $^{16}O$, to take two examples,
provide us with opportunities to  test the nuclear dynamical models, useful in
interpreting the  neutrino reactions, for example, those studied by the KARMEN
\cite{6} and the LSND \cite{7} collaborations. The latter have bearings on the
question of the neutrino oscillations, thus,
physics beyond the standard model. The inclusive NMC rates in
these targets are important benchmarks for the theory of  nuclear response,
of great value in setting the scale of our understanding
within  the standard model. While inclusive NMC rates are generally difficult
theoretically to compute \cite{8,9,10,10bis,2}, these are experimentally
relatively
easy to measure \cite{11}, particularly in muon factories, where high  muon
purity and a decent stopping rate are readily available \cite{12}.
Regrettably, such experiments have been few and far between in the
recent times \cite{11}. Our theoretical work here hopes to
elicit new  experimental interest in this subject, of importance
to nuclear physics and  particle physics.

The reaction of our interest is

\begin{equation}
\mu^- (1 S) + (A,Z) \rightarrow \nu_\mu + (A, Z-1)^*,
\end{equation}

\noindent
wherein the nucleus $(A,Z-1)^*$, formed in the reaction, is not
experimentally detected. The rate $\Lambda_c$ of the inclusive reaction
is experimentally inferred from the disappearance rate of the muon \cite{2,11}.

Theoretically, three common ways of calculating $\Lambda_c$ are: (1)
by computing and summing  exclusive channels \cite{10},  (2) by
using the sum rules \cite{8} and (3) by
using the elementary particle method \cite{10bis}.
 We follow a different method \cite{9}, which 
assumes the reaction (1) proceeding as

\begin{equation}
\mu^- (1S) + [p] \rightarrow \nu_\mu + [n],
\end{equation}

\noindent
where $[p]$ and $[n]$ are nucleons belonging to two different local
Fermi seas, characterized by the Fermi momenta $k_{F,i} (r)$
$(i = p,n)$, given by the medium densities $\rho_{i} (r) = k^3_{F,i} (r)
/3\pi^2$. Then, the differential rate for the reaction (2) is given,
in the local density approximation, by \cite{9}

\begin{equation}
\frac{d \Lambda_c}{d E_\nu} \sim \bar{\sum} \sum |T|^2 Im \, \bar{U}
(q_0, \vec{q}),
\end{equation}

\noindent
where we suppress in Eq. (3) phase space factors and measures of integration, shown fully in Ref. [9]; 
$|T|^2$ is the square modulus of the transition amplitude, suitably
renormalized \cite{9} in the nuclear medium, summed and
averaged over the initial and final states respectively; the function 
$\bar{U} (q_0, \vec{q})$ is the nuclear Lindhard function for the particle-hole
excitation. The three-momentum transfer $\vec{q}$ is fixed by the outgoing
neutrino momentum.  In our approach the  energy transfer, $q_0$,
to the nucleus  is given by

\begin{equation}
q_0 = E_\mu + Q' - Q - E_\nu.
\end{equation}

\noindent
Here $E_\mu$ is the available energy from the muon

$$
E_\mu = m_\mu - E_\mu^{1S},
\eqno{(5a)}
$$
where  the second term on the right-hand  side of (5a) 
is the muon binding energy.
$Q'$ comes into play in the nucleus with $N \neq Z$,

$$
Q' = E_{n_F}  - E_{p_F} \, ;
\eqno{(5b)}
$$
$Q$ is the threshold for the nuclear reaction (1) to begin \cite{13}

$$
Q = M_f - M_i \, ,
\eqno{(5c)}
$$

\noindent
the nuclear mass difference for the target and daughter nuclei. For light
nuclei under consideration in this Letter (target nuclei between $^6 Li$ and
$^{18} O)$, the $Q$ value varies from a low of
$0.67 \, MeV$, in the case of $^{14} N \stackrel{\mu^-}{\rightarrow}
\,  ^{14}C$ transitions, to a high of 21.15 MeV, in the case of
the  $^{14} C$ target, metastable, but still experimentally accessible, because of its
very  long life, leading to
$^{14} C \stackrel{\mu^-}{\rightarrow}
\,  ^{14}B$ transitions; indeed, $Q$ values exceeding $10 \, MeV$ are
quite common in light nuclei. We
find  a 
significant effect  of the $Q$ value on the total
capture rate in light nuclei, since the right-hand side of
(3) is a sensitive function of the available energy in the
reaction, (2).

The calculations in \cite{10} use a RPA approach to nuclear
structure. Kolbe {\it et al.} use a continuum RPA which sums
up over  the excited nuclear states above nucleon
emission. Auerbach {\ et al.} use a standard RPA, including pairing, 
that allows one to include
the contribution of all final states. In both cases the RPA evaluation
reduces the results with respect to the calculation with the
single particle 
orbitals, but the reduction is stronger in the
calculations of Auerbach {\it et al}.

Our approach 
might look simplified with respect to 
the ones just mentioned, but in fact
it is also an RPA approach built up from single particle states of an
uncorrelated local Fermi sea. This method 
in practice
is a very accurate tool when the excitation  energy is sufficiently large such
that relatively many excited states contribute to the process, and in 
particular if a large fraction of it comes from excitation to the 
continuum, 
as it is the case in $\mu^-$ capture. The adaptation of the method
to finite nuclei via the local density approximation has proved to
be a rather precise
technique to deal with
inclusive  photonuclear reactions \cite{CARR92}, response functions
in electron scattering \cite{GIL97} and deep inelastic scattering 
\cite{MARCO96}.

Obviously, because of its nature, the method only applies to inclusive
processes, summing over relatively many final states and it is
not meant to evaluate transitions to discrete states.

 The simplicity of the method, however, allows for improvements over the
traditional methods: 1) the sum $\bar{\Sigma} \Sigma |T^2|$ is done
relativistically; 2) the nucleon momentum of the Fermi sea is
considered in the calculations which usually   is taken equal to zero; 3) the
$Z_{eff}$ approximation is avoided and instead our evaluation uses
directly the muon wave functions. Nonrelativistic muon wave functions
are used, but
the relativistic effects were evaluated in \cite{9} and amounted to
about 2$\%$. The $\mu^-$ binding energy is also considered. 4) In
addition,   our approach builds
up the RPA correlations by allowing the $\Delta h$
excitation on top of the $ph$ excitation considered in \cite{10}.
This is a {\it major difference} with the 
other theoretical RPA  approaches mentioned \cite{10}. A Landau-Migdal 
interaction is used to account for the propagation of these $ph$ and 
$\Delta h$ states in the nucleus.

The $\Delta h$ excitation leads
 to extra quenching of the transition strength
with respect to the one found by Kolbe {\it et al}. The role of this 
$\Delta h$ excitation in the quenching of Gamow Teller transitions
was emphasized by  Rho and later by Brown and Rho in ref \cite{4}.   
This quenching could be 
interpreted as an effective quenching of $g_A$, since this 
coupling term gives
the largest contribution to the $\mu$ capture rate; 
we can, instead,
talk in terms of quenching of the transition strength. The pseudoscalar 
current is also renormalized, but this term plays a smaller role in
$\mu$ capture. 

Our method was used in \cite{9}, where a good overall agreement with 
experiments over  the periodic table was found. 
The method has proved
even more accurate than 
anticipated in the ref. \cite{9}, when the
improvements of the present paper
are done.
The basic framework of our approach has  been described in detail
in \cite{9}. We stress here only significant improvements of this
approach: (1) We use a new Lindhard function \cite{14}, which takes into
account a shift of the neutron and proton Fermi seas due to 
the experimental Q value \cite{13}. Its imaginary part coincides with the
the results of the Eq.(3) in terms of the ordinary Lindhard function,
but its real part is changed slightly with respect to the old one. The
new Lindhard function avoids pathologies of the ordinary
Lindhard function in the limit
($q_0, \vec{q}) \rightarrow 0$. This  limit, however, does not occur in
the muon capture kinematics, but it does play an important role
in the neutrino scattering at low energies. (2) The experimental $Q$
value and the theoretical one, $Q'$, are used in the evaluation
as indicated in eqs. (3), (4).
The Q-value in general is important
in determining the NMC rate in light nuclei.
(3) The uncertainties coming from the Landau-Migdal parameter $g'$
are included by varying $g'$ in the range \cite{15}

$$
g' = 0.7 \pm 0.1 \, .
\eqno{(6)}
$$
This is an important parameter influencing the nuclear
response in our approach. This parameter occurs in the particle-hole and the
Delta-hole interactions \cite{4,5}.
(4) In many nuclei, the radial form 
of the nuclear density
has parametric uncertainties. We take them  also into account in
our calculation by letting the density parameters vary  within the
experimental errors of \cite{16}.

In  Table I, we display our calculated total capture rate $\Lambda_c$,
obtained by integrating (3), for $g' = 0.7$ and for one set of
standard nuclear density parameters \cite{16}. 
The present approach has 
significantly improved the agreement between the theory and
experiment, compared with  results 
reported in \cite{9}.

The principal uncertainty of our theoretical estimate of the NMC rate
stems from that of the parameter $g'$. For the range of $g'$, given in 
equation (6), this results in an uncertainty of about $\pm(5-7)\%$ in the
 NMC rate for all the nuclei considered in this paper and reported in
table I. 
 Thus, to take a physical example, that of
NMC by $^{12}C$, we vary $g'$
from 0.6 to 0.8 and this results in a variation of the rate from
$3.78 \times 10^4 \, s^{-1}$ to $3.43 \times 10^4 \,
s^{-1}$, a $\pm 5 \%$ effect 
around the central value of $3.6 \times 10^4 \, s^{-1}$, corresponding
to $g' = 0.7$.  Our experience with many
electroweak and strong processes points towards a value of $g' = 0.7$
\cite{CARR92,GIL97,MARCO96}. 
Similarly, other implicit parameters of the particle-hole or the
Delta-hole interactions are determined from other processes. 

The nuclear densities $\rho_{p,n}$, enter in our calculation via the local
density approximation \cite{9}. There are three cases, for A equal to 12,
14 and 16, in which radial uncertainties are readily investigated
in our approach by calculating the capture rate for
many sets of density parameters determined from electron scattering
\cite{16}. They produce relatively small uncertainties of $\pm(2-3)\%$.
The theoretical capture rates quoted for these nuclei in table 1 should 
be read with this additional uncertainty due to the nuclear densities.
Thus, in $^{12} C$, the total
capture rate varies from $3.45 \times 10^4 \, s^{-1}$ to $3.60\times
10^4 \, s^{-1}$, keeping the spin-isospin parameter fixed. This is
a $\pm 2$ $\%$ variation due to uncertainty of the nuclear
density around the central value. 
Assuming the uncertainties 
of $g'$ and nuclear densities
to be independent, we have a theoretical uncertainty in $^{12}C$ NMC
rate of $\simeq 6 \%$: this is a conservative estimate of our
theoretical error. Thus, the $^{12}C$ NMC rate is, according to our
theoretical calculation,

$$
\Lambda_c (^{12}C) = (3.60 \pm 0.22) \times 10^4 \, s^{-1}\, .
\eqno{(7)}
$$

This is in good agreement with the best experimental determination
so far \cite{11}, taking the world average of the best determinations,
with their errors in quadrature:

$$
\Lambda_{c}^{exp} (^{12} C) = (3.80 \pm 0.10) \times 10^4 \, s^{-1} \, .
\eqno{(8)}
$$

 In the following, we  compare our work with the recent work of
Kolbe {\it et al.} and Auerbach {\it et al.} for this reaction.

Kolbe {\it et al.} \cite{10} have done a continuum RPA calculation, which can
 compute with a reasonable accuracy the NMC rates to the 
particle-unbound  states
in $^{12}B$. 
Using their calculation of the 1994 paper [10], we obtain the inclusive
capture rates for the $^{12}C$ nucleus, wherein the particle bound states 
are not included for the excited nucleus (see their Table I). We get 
theoretical rates for two different potentials to be 3.42 and 3.34 in units 
of $10^4 s^{-1}$. These numbers are slightly above  the experimental value
for the capture to the particle unbound states quoted by Kolbe {\it et al.}
Taking a conservative lower limit
of $7000 \; s^{-1}$ for the muon capture rate to the 
particle-bound states directly from the
experiment, we can translate the calculation of Kolbe {\it et al.} as

$$
\Lambda_c (^{12}C) \simeq ( 4.04 - 4.12 ) \times 10^4 \, s^{-1}. \,
\eqno{(9)}
$$
Auerbach {\it et al.} have several variations of their HF-RPA model and they
obtain \cite{10}

$$
\Lambda_c (^{12}C) \simeq (3.09 - 3.64) \times 10^{4} \, s^{-1}.
\eqno{(10)}
$$
 
Recent researches in neutrino physics have brought NMC in $^{12}C$ in
sharper focus as a theoretical benchmark. One has to keep in mind the
results of the recent $(\nu_e, e^-)$ and $(\nu_\mu, \mu^-)$ experiments
\cite{6,7}, along with the results
of some recent theoretical calculations\cite{10,18}. The latter reaction  
is the {\it inverse} of the NMC. In particular, the recent LSND 
experiment has produced an
inclusive $(\nu_\mu, \mu^-)$ rate, which is {\it in
disagreement} with the calculation of Kolbe {\it et al.} \cite{10} by a
factor of about 1.5. However, Auerbach {\it et al.} \cite{10}, in a recent
paper find that this disagreement can be eliminated in their Hartree-Fock-
Random-Phase Approximation (HF-RPA) calculation that includes 
pairing. With the present approach, we obtain a rate which is within
15 $\%$ of the experimental result when the maximum uncertainties in the
theoretical as well as experimental results are taken into account\cite{18}.
 
In the neutrino physics context, the NMC produces
the following situation: 
 while Kolbe {\it et al.} \cite{10} obtain a result for NMC rate 
which is within $5\%$ of the experimental result and explain the  
$(\nu_e,e^-)$ inclusive cross sections rather well, they overestimate 
the inclusive $(\nu_{\mu}, \mu ^-)$ cross section by (45-50) $\%$.
In the calculations of Auerbach {\it et al.} \cite{10}, it is possible 
in one version of their model to obtain a NMC rate 
$(\Lambda _c =3.64 \times 10^4  s ^{-1})$ in fair agreement with the 
experiment, but this version of the model predicts slightly
higher values for  the $(\nu_{\mu}, \mu ^-)$ inclusive cross sections 
and a large value for the $(\nu_e,e^-)$ inclusive cross sections when
compared with the experiments .
On the other hand, the version of the model, which can explain the 
inclusive neutrino reactions , predicts a NMC rate of
 $3.09 \times 10 ^4 s^{-1}$ which is  smaller than the experimental value. 
 Our present method, when applied to $^{12}C$
reproduces  the experimental results of muon capture as well as
those of the  
$(\nu_e,e^-)$ inclusive cross sections quite well, but it also 
overestimates  the experimental values for  the $(\nu_{\mu}, \mu ^-)$
inclusive cross sections by about (15-20)$\%$ \cite{18}.
This  underscores {\it the important role  of the NMC} in selected
targets  as benchmark reactions to calibrate
the nuclear theory, which is, in turn, used to delineate standard model
physics. From this point of view, the NMC rates in $^{12,13}C$ and
$^{16}O$ are particularly important.

The above discussion, comparing results of different accurate methods,
gives us an idea of present theoretical uncertainties in calculations of 
weak nuclear reactions.

We now come to a very subtle physics in NMC, the isotope effect. In Table II,
we display our calculations for the
ratios of the rates of NMC for a pair of isotopes.
We compare here our results with experimental ones wherever available, and
the prediction of the well-known  formula of Primakoff
\cite{8}:

$$
\Lambda_c (A', Z')/ \Lambda_c (A,Z) =
\frac{1 - \delta (A' - Z')/2A'}{1 - \delta (A - Z)/2 A}.
\eqno{(11)}
$$

\noindent
where $\delta \simeq 3.15$. The
Primakoff formula yields Pauli blocking of the muon capture process
rather approximately.

From  Table II, we can see that our
calculation does much better than the Primakoff formula in three out of four cases
where there are data on isotope shifts. For
example, our calculation yields for $^{13} C/^{12}C$ and $^{18}O/^{16}O$ ratios
0.82 and 0.78, in fair agreement with the experimental values of
$0.90 \pm 0.01$ and $
0.86 \pm 0.15$ respectively, in contrast to the
predictions of the Primakoff formula of 0.71 and 0.59 in the two cases.
Our approach is able
 to take into account the special properties of individual
target nuclei such as the shell closure in $^{16} O$ and the large $Q$-value
needed in $^{14}C$. No such physics is included in the relatively
simple formula (11).

Detailed calculations, 
such as those in \cite{10}, should likewise
do better than the Primakoff  formula.
They also take into account the role of the
experimental $Q$ values. The case of $^7 Li/ ^6 Li$ does not
come up too good in our approach.
This is understandable,
since, in 
very light nuclei, the local density 
approximation  can become 
less accurate.

We now make some remarks on the experimental situation and future prospects
at the muon factories. It is important to have muon capture rates in light
nuclei measured with high precision. In several targets,
old \cite{2} and newer \cite{11} experiments disagree quite strongly
(e.g. $^{9} Be, ^{14}N, ^{16}O$ etc.), even though the more recent experiments
are of higher statistical accuracy. While our theoretical estimates agree with
most of the recent determinations of the $\Lambda_c$, there are
significant disagreements in $^{7}Li$ and $^{14}N$. Some targets, $^6 Li$
and $^7 Li$, have relatively poorly known
$\Lambda_c$ and deserve better experimental determination. $^{14}C$, being a
radioactive nucleus, will pose experimental challenges
to prepare a dense enough nuclear target.
However, it is very interesting theoretically: the $Q$-value
for NMC in this target is the largest in the nuclei studied here, with
an enormous effect on the rate. Hopefully, the long half-life
of $^{14}C$ would make an experiment with it feasible.

Our worst agreement with experiment comes in $^7 Li$ and $^{14}N$
targets. We predict the inclusive NMC rates $3.4 \times 10^3$   and $8.7
\times 10^4 \, s^{-1}$ in these two cases, while the most accurate experiments
give $(2.26 \pm 0.12) \times 10^3 \, s^{-1}$ and
$(6.93 \pm 0.008) \times 10^4 \, s^{-1}$ respectively \cite{11}. Given the
quality of our agreement in other targets with experimental
results,  this disagreement invites an
experimental reconfirmation, with a special care on the systematic
errors of the experiments.
Further experimental studies are also needed to investigate possible 
non-statisticality in the hyperfine states of the muonic atom \cite{2},
in particular, in the case of $^{14}N$, due to hyperfine conversion. 

In summary, we have studied total NMC rates in the framework of
a theory  wherein vector and axial-vector
strengths in nuclei are appreciably  renormalized due to particle-hole
and Delta-hole correlations. Large effects  from
the consideration of the Q-values of
the NMC are seen,
in some nuclei. A reasonable ($\pm$ 0.1) variation of the Landau-Migdal
parameter $g'$ around its preferred value 0.7 translates into a 
$\pm (6-7) \%$
uncertainty of the NMC rate for the nuclei studied in this paper. 
Similarly the nuclear radial uncertainties in some of these nuclei 
contribute  at most a $\pm (2-3) \%$ variation of the capture rate.

Since renormalizations of weak vector and axial-vector strengths, studied here, also
occur in the electromagnetic and strong processes, we should interpret them
as those for nuclear strengths rather than for weak
nucleon couplings in nuclei. This has important
consequences in our understanding of the nuclear QCD effects,
in particular, the Bjorken sum rule in nuclei.

Precise theoretical understanding of the NMC in $^{12}C$,  $^{13}C$
and $^{16}O$ would provide benchmarks in low-energy neutrino
physics, with important bearings on the issue of neutrino oscillations 
and physics beyond
the standard model. Although small, there are still 
defferences between the predictions of different
accurate models and it would be most desirable to 
examine and test  these
differences in precise experiments in  future.

\vspace{0.4cm}

We thank Prof. J. Deutsch  and Prof. D. Measday for valuable communications.
Three of us (HCC, NCM and SKS) have the great pleasure of thanking
Professor E. Oset for his warm hospitality at the Universidad de Valencia. One of us
(NCM) is grateful for the generous support of  ``IBERDROLA de Ciencia
y Tecnologia" and  acknowledges the partial support of the U.S. Dept. of Energy.
HCC and SKS acknowledge the support of the Ministerio de Educacion y
Cultura of Spain in their sabbatical stays. This work is also
supported in part by CICYT, Contract numbered AEN-96-1719.

\vspace{0.4cm}

{\bf Table I}: Total muon capture rates in $s^{-1}$ in various
target nuclei, 
compared with the most accurate experimental results cited in the
literature, wherever available\cite{11}; A dash
in the last column means that no data 
are available in that case;
in case of two or more experimental results of comparable quality, we
display the value with best precision here. We use $g' = 0.7$ and one 
standard set  of
radial densities \cite{16}.

\begin{center}
\begin{tabular}{llc}
Target &  This calculation  
& Experiment\\
$^6 Li$ &  $4.68 \times 10^3$ &
$(4.68 \pm 0.12) \times 10^3$\\
$^7 Li$ &  $3.41 \times 10^3$ &
$(2.26 \pm 0.12) \times 10^3$\\
$^9 Be$ &  $8.84 \times 10^3$ &
$(7.4 \pm 0.5) \times 10^3$\\
$^{10} B$ &  $2.67 \times 10^4$ &
$(2.78 \pm 0.07) \times 10^4$\\
$^{11} B$ & $1.86 \times 10^4$ &
$(2.19 \pm 0.07) \times 10^4$\\
$^{12} C$ &  $3.60 \times 10^4$ &
$(3.76 \pm 0.04) \times 10^4$\\
$^{13} C$ &  $2.95 \times 10^4$ &
$(3.38 \pm 0.04) \times 10^4$\\
$^{14} C$ &  $2.34 \times 10^4$ &
$-$\\
$^{14} N$ &  $8.67 \times 10^4$ &
$(6.93 \pm 0.008) \times 10^4$\\
$^{15} N$ & $6.34 \times 10^4$ &
$-$ \\
$^{16} O$ &  $1.16 \times 10^5$ &
$(1.026 \pm 0.006) \times 10^5$\\
$^{17} O$ &  $1.06 \times 10^5$ &
$-$ \\
$^{18} O$ & $9.00 \times 10^4$ &
$(8.80 \pm 0.15) \times 10^4$\\
\end{tabular}
\end{center}

\vspace{0.5cm}

\vspace{0.5cm}

{\bf Table II}:  Isotope effect (with $g' = 0.7)$ calculated by us,
compared with the Primakoff formula (Eq. (14)) and experiment\cite{11}.

\begin{center}
\begin{tabular}{llll}
Isotope pair & This calculation & Primakoff & Exp. \\
$^7 Li /^6 Li$ & 0.73 & 0.47 & 0.48 $\pm$ 0.03\\
$^{11} B /^{10} B$ & 0.79 & 0.67 & 0.79 $\pm$ 0.03\\
$^{13} C /^{12} C$ & 0.82 & 0.71 & 0.90 $\pm$ 0.01\\
$^{14} C /^{12} C$ & 0.65 & 0.47 & -\\
$^{14} C /^{13} C$ & 0.79 & 0.66 & -\\
$^{15} N /^{14} N$ & 0.73 & 0.75 & -\\
$^{17} O /^{16} O$ & 0.91 & 0.78 & -\\
$^{18} O /^{16} O$ & 0.78 & 0.59 & 0.86 $\pm$ 0.15\\
$^{18} O /^{17} O$ & 0.85 & 0.75 &  -\\
\end{tabular}
\end{center}

\newpage

\end{document}